\documentstyle[12pt,epsf]{article}
\begin{document}
\baselineskip 16pt plus 2pt minus 2pt

\hfill TK 96 04


\vspace{1.0cm}

\begin{center}

{\large { \bf Kaon loops in pion photoproduction: Facts and fancy}}
\vspace{1.0cm}

Sven Steininger and Ulf-G. Mei\ss ner

\vspace{0.8cm}

Universit\"at Bonn, Institut f\"ur Theoretische Kernphysik, Nussallee
14-16,\\ D--53115 Bonn, Germany

\vspace{0.4cm}

\end{center}

\begin{abstract}
\noindent We show that a recent claim of huge kaon loop corrections to
the electric dipole amplitude in neutral pion photoprodcution at
threshold is incorrect. The difference between the two and three
flavor chiral perturbation theory calculation is marginal and thus
previous claims of a good understanding of this reaction remain valid.
\end{abstract}

\vspace{1.0cm}

\noindent In a recent paper, Banerjee and Milana \cite{bmjunk} claim
that there are huge kaon loop corrections to the electric dipole
amplitude $E_{0+}$ for neutral pion photoproduction at threshold. If
that were true, all previous calculations of Bernard et al. should be
considered incomplete and the reaction $\gamma p \to \pi^0 p$ could not
be used to  test the chiral dynamics of QCD (for a review, see \cite{bkmr}). 
While there are still
some open questions concerning the convergence in the two--flavor
case, the recent analysis of the new TAPS and SAL data does indeed lend credit
to the approach and its extension to pion electroproduction
\cite{bkmprl,bkmprl2}. As spelled out in detail in Ref.\cite{bkmz}, in the
threshold region one has to consider two basic Feynman graphs, the 
so--called triangle and rescattering diagrams (plus their crossed
partners). Evaluating these correctly with intermediate $\Lambda,
\Sigma^0$ states, one finds the following kaon loop contribution for 
$\gamma p \to \pi^0 p$ at order $q^3$ in the chiral counting :
\begin{equation}
E_{0+}^{q^3,K} (\omega) = \frac{e\, F}{64 \pi^2 F_\pi^3} \biggl( M_K^2 \,
\arcsin \frac{\omega}{M_K} -\omega \, \sqrt{M_K^2 - \omega^2} 
\biggr) \, \, ,
\label{E0correct}
\end{equation}
with $\omega $ the pion energy in the $\pi N$ cms frame and all other
quantities are standard. We point out, however, that we  could
have used here $F_K = 1.2 F_\pi$ as well. This would reduce this contribution 
considerably. The difference to the result in \cite{bmjunk} is the
factor in front of the square root in Eq.(\ref{E0correct}). This can
be traced back to the fact that Eq.(13) in \cite{bmjunk} 
has the wrong sign, see
e.g. ref.\cite{bkmz}. A detailed account of the SU(3) extension of the
work by Bernard et al. for SU(2) is given in Ref.\cite{sven}. With
$F=0.5$, $F_\pi =93\,$MeV, $e^2/4 \pi = 1/136.037$ and $M_K = 495 \,$
MeV, one has at threshold
\begin{equation}
E_{0+}^{q^3,K} (\omega= M_{\pi^0}) =  0.14 \cdot 10^{-3}/ M_{\pi^+}
\, \, \,  \, \, ,
\label{E0val}
\end{equation}
which is well within the theoretical uncertainty of the SU(2) calculation
and can be accounted for by minor adjustments of the low energy
constants $a_3$ and $a_4$, see \cite{bkmprl2,bkmz}. That the same holds
for the neutron amplitude does not need to be elaborated on here.
It is important to stress that in the SU(2) calculation with a fixed,
non-zero strange quark mass, the kaons are frozen and can only
contribute at order $M_\pi^3$. To be precise, one finds
\begin{equation}
E_{0+}^{q^3,K} (\omega=M_{\pi^0}) = \frac{e\, F \, M_\pi^3}
{96 \pi^2 F_\pi^3 \, M_K}
\label{E0frozen}
\end{equation}
which leads to almost the same result as in Eq.(\ref{E0val}).
Note that Eq.(14) given in \cite{bmjunk} in the limit of fixed
strange quark mass violates even the current algebra result 
for $E_{0+}$ at ${\cal O}(M_\pi)$.

Finally, we would like to point out that the importance of measuring
the electric dipole amplitude for the neutron has already been lucidly
discussed in the 1992 paper by Bernard et al. \cite{bkmnpb} and
stressed again in \cite{bkmz}. The reason to measure this quantity is
to get a handle on possible
isospin violations in the pion--nucleon system and not because of the
suppression of kaon loop effects in the difference $E_{0+}^{\pi^0 p} -
E_{0+}^{\pi^0 n}$ as claimed by Banerjee and Milana.

\vskip 1cm

\noindent We thank V\'eronique Bernard and Norbert Kaiser for comments and
checks.

\vskip 1cm


\end{document}